# Revolutionising Role-Playing Games with ChatGPT

**Rita Stampfl**  rita.stampfl@fh-burgenland.at
*Department Information Technology*
*University of Applied Sciences Burgenland*
*Eisenstadt, Austria*

**Barbara Geyer**  barbara.geyer@fh-burgenland.at
*Department Information Technology*
*University of Applied Sciences Burgenland*
*Eisenstadt, Austria*

**Marie Deissl-O'Meara**  marie.deissl-omeara@fh-burgenland.at
*Department Information Technology*
*University of Applied Sciences Burgenland*
*Eisenstadt, Austria*

**Igor Ivkic**  i.ivkic@lancaster.ac.uk
*Department Computing and Communications*
*Lancaster University, Lancaster*
*United Kingdom*

**Corresponding Author:** Rita Stampfl

## Abstract

Digitalisation in education and its influence on teaching methods is the focus of this study, which examines the use of ChatGPT in a role-playing game used in the Cloud Computing Engineering Master's programme at the University of Applied Sciences Burgenland. The aim of the study was to analyse the impact of AI-based simulations on students' learning experience. Based on Vygotsky's sociocultural theory, ChatGPT was used to give students a deeper understanding of strategic decision-making processes in simulated business scenarios. The methodological approach included role-playing and qualitative content analysis of 20 student reflections. The findings suggest that ChatGPT enhances students' engagement, critical thinking, and communication skills, in addition to contributing to the effective application of theoretical knowledge. Furthermore, simulations can contribute to the effective application of theoretical knowledge. The results underscore the significance of adaptive teaching approaches in promoting digital literacy and equipping learners for the digital workplace. The integration of AI into curricula and the need for ongoing innovation in higher education are also emphasised as a means of guaranteeing excellent, future-focused instruction. The findings highlight the potential of AI and ChatGPT in particular, as an innovative cutting-edge educational tool that can both enhance the learning experience and help achieve the Sustainable Development Goals (SDGs) through education.

**Keywords:** Artificial intelligence, ChatGPT, Digital education, Role-playing games, Simulation games.

## 1. INTRODUCTION

The COVID-19 pandemic has accelerated the digital transformation of education and highlighted the need to adapt teaching methods to the digital age [1]. This sudden change has emphasised the importance of strategic planning to improve curricula and make digital learning more engaging [2]. It also encompasses the difficulty faced by educators of fusing the best technology with the best human teaching to ensure future-ready education [3]. Making content relevant to students' lives and providing opportunities for them to interact in real-world learning situations with other qualified adults and students are at the core of future-ready education. Students in future-ready education should be taught how to apply what they have learned to ensure that they can effectively meet the needs of the future job market.

Tran and Herzig [4], found that innovative course design greatly improves students' learning outcomes in addition to producing positive learning experiences. Students themselves express their appreciation for the complementarity and interaction of learning activities, surrounding both asynchronous and synchronous sessions. Furthermore, the use of digital learning tools offers a unique opportunity to promote intense engagement through their broad reach while creating personalised learning environments tailored to the needs and knowledge of the individual. Thus, focused and productive learning experiences are made possible [5].

Against this background, this study explores the use of ChatGPT in business games in the Master's programme in Cloud Computing Engineering as an innovative method to enhance the learning experience. By incorporating AI, the method not only increases student motivation and engagement but also promotes the development of key skills such as critical thinking and problem-solving. In the realm of education, leveraging the potential of AI not only aligns with the evolving pedagogical landscape but also serves as a powerful means to address the advancement of Sustainable Development Goals. Thus, high-quality higher education and equitable access to it are critical to achieving these goals, and even constitute a goal in their own right under SDG 4. The challenge is to fully exploit the potential of AI in education to improve learning outcomes and prepare learners for a digitised future. At the same time, the associated risks need to be carefully managed, as recommended by the European Commission [6].

## 2. THEORETICAL BACKGROUND

The responsible use of Artificial Intelligence (AI) requires active integration into teaching, ensuring alignment with educational objectives. AI is capable of performing a wide range of functions in higher education [7]. De Witt et al. [8], point out that "in the future, AI tools [...] can also be used to train metacognitive skills such as critical and creative thinking, reasoning, planning, decision making and problem-solving". A comprehensive study by Abrami et al. [9], found that the most successful strategy used a combination of subject-based and cross-curricular interventions. AI tools are already widely used in communication-focused professions [10], and are predicted to quickly expand to other professions. Thus, the competent and reflective use of AI tools has gained increased importance in light of universities' obligations as part of a future skills framework to ensure that students are equipped to cope with a rapidly changing environment and to prepare them for the future labour market. This is described by Ehler [11], as an essential component of digital literacy. In a didactic context, the following areas of application are discussed under the heading of 'artificial intelligence technologies' [12]:

- Intelligent Tutoring Systems (ITS)
- Natural language processing, automatic speech recognition and automatic text generation (NLP, ASR, NLG)
- Automatic assessment and grading
- Multimedia human-machine interaction (e.g. chatbots, learning companions)
- Learning (predictive) analytics, data mining in the education sector (LA, LPA, EDM)
- Adaptive learning, recommendation services

AI-based tools can be used both in administrative areas and in the teaching process itself. On the one hand, they can support the organisation of studies, and on the other, they can help make teaching and learning environments more flexible, personalised, and efficient [7].

The use of AI tools such as ChatGPT in education reflects the principles of various educational theories, particularly Vygotsky's sociocultural theory. This theory stresses the importance of social interaction in cognitive development, with learning being understood as a deeply rooted social process [13]. Recognising the significance of this sociocultural perspective, ChatGPT promotes this understanding by simulating social interaction and providing access to a wide range of cultural knowledge, creating a learning environment that mimics real-life sociocultural contexts. This is in keeping with Vygotsky's theory whose assumption it is that people learn through their experiences in the Zone of Proximal Development (ZPD) [14], and which emphasises language-based learning.

In the light of these developments, it is imperative to comprehend the features of the contemporary student population at universities of applied sciences. Statistics Austria [15], reports that in the winter semester of 2022/23, 96.33% of students at such institutions in Austria were under 40 years of age. These students born after 1980 have been shaped by the introduction of computers into many areas of life [16]. They have grown up in media environments with videos, consoles, and computer games [17], which involve the players in entertaining activities. In the same vein, a simulation game which is a unique and creative tool can be used to increase students' interest in learning [18]. Simulation game learning is based on theories and applications that aim to improve student engagement and performance by involving them in real-life situations [19, 20]. By creating immersive and interactive learning experiences, simulation games effectively bridge the gap between theoretical knowledge and practical application, enhancing the overall educational impact. Through engaging in these games, students can develop their ability to comprehend things and evaluate their expertise. They can develop their ability to analyse given information, express their own point of view, formulate thoughts clearly, foresee possible consequences of different solutions, and compare their own level of understanding and perception with others. Simulation games increase interest in learning and help teachers to motivate students [21], and they can provide both learners and teachers with tools that facilitate active problem solving. They are seen as a practical approach to knowledge acquisition [18], and can also be used in the field of university teaching [22].

Role-playing can play an important role in the learning process with research showing that role-play can positively influence different aspects of learning, including cognitive performance, motivation and emotional engagement [23–25]. For example, role-playing combined with personalised learning has been shown to be an effective tool for improving student learning, although integrating personalised learning into role-playing has been challenging [26]. Furthermore, role-playing games are said to have transformative potential that benefits individuals in psychological, social, educational, therapeutic, professional, and political dimensions [27].

Elaborating on this transformative element, the third level of Bloom's taxonomy of learning objectives [28], entails applying knowledge practically. The act of practicing involves applying what has been learned to new and concrete situations, and the dynamic and interactive aspects of role-playing games greatly facilitate this process. The idea of actively integrating an AI tool in a business game stemmed from observations made by Schmid et al. [12], who cited AI tools for use in multimedia human-machine interaction using chatbots as an example. Additionally, Matute Vallejo and Melero [18], viewed business games as active problem solving and integration into real-life situations. Hence, the idea arose to actively use an AI tool in a business game for students to apply what they have learned in a concrete situation. This method encourages the development of skills and practical understanding by allowing students to apply their learned material in a real-world setting. Thus, the AI tool ChatGPT is actively used as a chatbot to simulate conversational situations for students [29]. The benefit of this is that ChatGPT assumes the role of the interlocutor, allowing students to play an asynchronous simulation game at their own pace and without being restricted to a particular location or time.

## 3. METHODOLOGY

The study was conducted using the course Impact of Cloud Computing on Organisations from the Cloud Computing Engineering Master's programme. The overall aim of the course is for students to be able to recognise the impact of cloud technologies on international and multicultural organisations, and to apply the social skills they have learned in relation to change, negotiation, and decision-making. To test these social skills, a task was designed that required students to apply the technical and application-oriented content taught in the course through role-playing in a fictitious negotiation using ChatGPT. As part of the course, students developed a project assignment for a cloud migration project. The assignment was based on the project brief and the benefits of cloud computing, as cited by Salesforce [30]. As part of the assignment, a specific prompt to start a role-play in ChatGPT was designed and made available to students. Students were asked to actively argue and negotiate the approval of the project budget in a role play. FIGURE 1 shows the prompt:

> Let's role play. You are the CEO of a company that is thinking about moving all its on-site hosted servers, including its services, to the cloud. However, you are not sure about this decision and are very critical of the cloud in general. I want to enter a project budget of **>>amount from your project order<<** Euro and you want to give it to me only after intense negotiations. You ask me critical questions about the following areas: Cost savings, security, flexibility, mobility, insight, increased collaboration, quality control, disaster recovery, loss prevention, automatic software updates, competitive advantage and sustainability. I am a cloud consultant who answers your questions. You ask one question at a time and ask the next question based on my answer. I only get a commitment from you if the IPMA project assignment criteria are met.

Figure 1: Prompt for Starting a Role-Playing Simulation Game using ChatGPT.

In ChatGPT 3.5 lecturer-led conversations can only be prompted; they cannot be altered once they have commenced. This implies that the conversation was different for each student. In addition to the role-play, the students also reflected on their subjective perception of the simulation game as part of this task.

### 3.1 Participants and Data Collection

As part of the case study, 20 students completed the required written reflection on the assignment. The students were a homogeneous group of prospective cloud specialists in the first semester of the Master's programme, all of whom already held a computer science bachelor's degree or above. The size of the sample was set at 20 participants and detailed information on their demographical background is given in TABLE 1.

In the students' written reflection students were asked to provide a 300–500 word written reflection in which they discussed their subjective perception of the task. The data analysis was based on the written reflections of the students. The data collection was conducted just once, in October and November of 2023.

### 3.2 Data Analysis

In order to systematically analyse the data collected from the written reflections on the case study task, a thorough content analysis was essential. As Kuckartz and Rädiker [31], explained, qualitative content analysis is a methodically controlled scientific analysis of texts, images, films, and other forms of communication. Qualitative content analysis comprises three basic approaches: content-structuring, evaluative, and typifying qualitative content analysis. Given the research design of this study, a content structuring approach was chosen. The data collection- and data analysis process is shown in FIGURE 2.

Table 1: demographical background of the participants

| Participant | Academic degree | Gender | Age | Citizenship |
|---|---|---|---|---|
| S1 | BSc | Male | 29 | Austria |
| S2 | BSc BSc MSc | Male | 28 | Austria |
| S3 | BSc | Male | 29 | Croatia |
| S4 | BSc | Male | 26 | Austria |
| S5 | BSc | Male | 33 | Austria |
| S6 | BSc | Male | 25 | Hungary |
| S7 | BA MA | Male | 31 | Austria |
| S8 | BSc | Male | 21 | Austria |
| S9 | BSc | Male | 27 | Austria |
| S10 | BSc | Male | 23 | Austria |
| S11 | BSc | Male | 43 | Austria |
| S12 | BSc | Male | 33 | Austria |
| S13 | BSc | Male | 40 | Austria |
| S14 | BSc | Male | 33 | Austria |
| S15 | Bakk. tech. | Male | 40 | Austria |
| S16 | BSc | Male | 29 | Austria |
| S17 | BSc | Male | 27 | Austria |
| S18 | BSc | Female | 40 | Austria |
| S19 | BSc | Male | 27 | Switzerland |
| S20 | BSc | Male | 43 | Austria |

Content structuring in qualitative content analysis methodically divides information into different categories [31]. This structured framework serves as the basis for a systematic and organised presentation of the research findings. The content analysis category system was developed deductively. The following overarching categories were derived directly from the case study: (1) simulation game, (2) ChatGPT, and (3) learning experience.

These overarching themes are closely related to the case study task and effectively summarise the focus of this research. In line with the research objective, the key themes identified revolve around simulation games using ChatGPT and the learning experience it provides. Furthermore, these basic categories serve as a solid foundation for the design and organisation of the findings. The process of delineating types through content analysis is greatly facilitated by the use of quality data analysis (QDA) software, a concept recommended by Kuckartz and Rädiker [31]. Consequently, the MAXQDA 2022 standard was carefully used to streamline the content analysis in this particular study.

In the coding phase, the collected data were thoroughly analysed in terms of the main thematic categories and systematically coded accordingly. Once the coding process was complete, a series of basic and complex analyses were carried out. The analyses were structured according to pre-defined categories. The content analysis of the reflections included both qualitative components and aspects such as verbatim quotes [31].

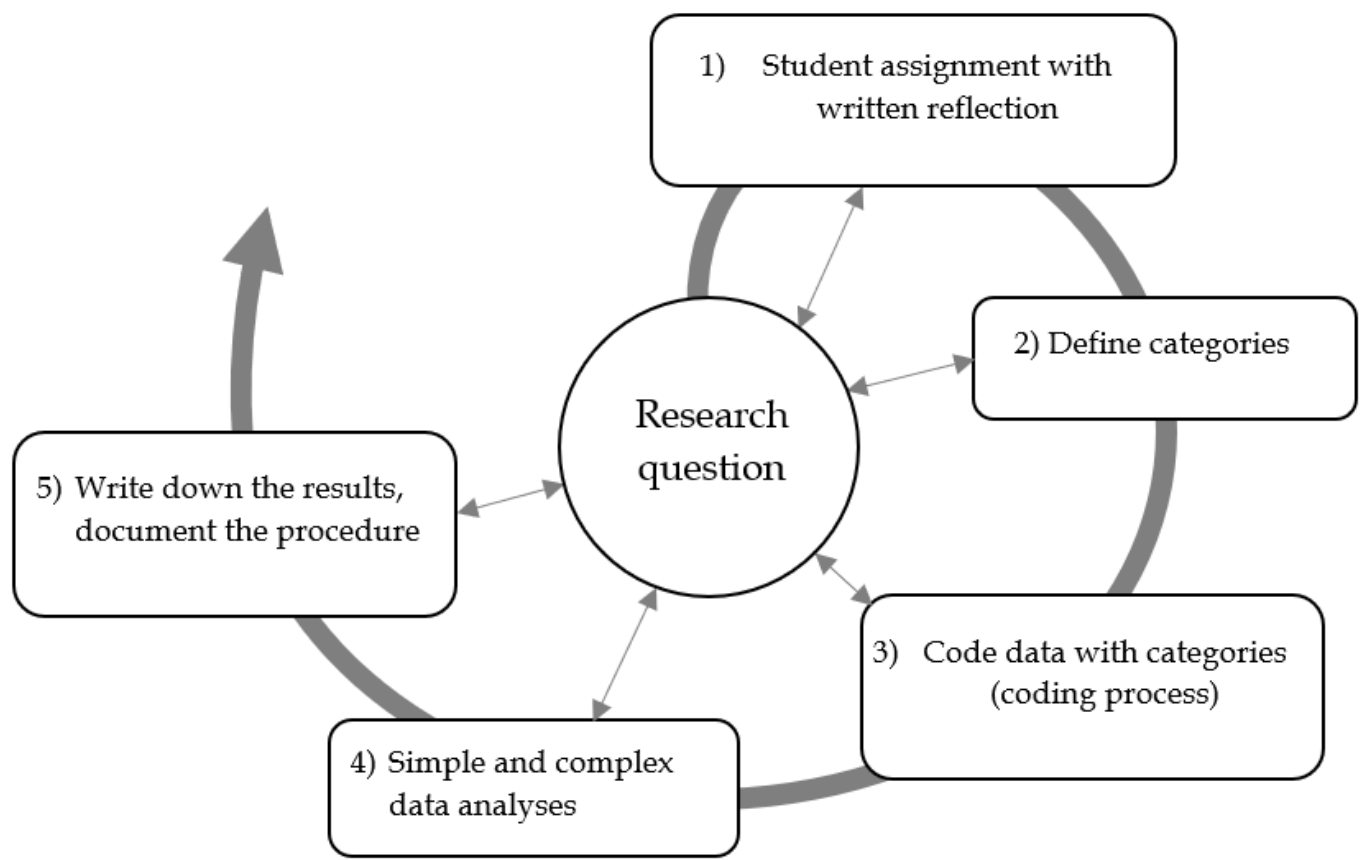

Figure 2: data collection and data analysis process based on Kuckartz and Rädiker [31]

## 4. RESULTS

The results are listed by category, and these categories represent the students' experience of the exercise, their insights into the functionality and effectiveness of ChatGPT in simulated role-playing business conversations, and their evaluation of the simulation as a learning method. Students report a valuable learning experience that provides deeper insights into the decision-making and strategic planning of cloud migration. They highlight the challenges and effectiveness of ChatGPT in simulating realistic conversational situations, although some point out the limitations of AI in terms of depth of interaction and the ability to continue a conversation naturally. The simulation is seen as an innovative and effective way to improve communication skills and understanding of cloud migration.

### 4.1 Business Game

The Business Game category summarises students' evaluations and reflections on the effectiveness of the simulated business game as a learning method. This category includes insights into the dynamics of the game, the role of interaction with the ChatGPT, and the overall evaluation of the simulation as a pedagogical tool.

Students described the simulation as an innovative and interactive way to understand complex cloud migration concepts and processes. The simulation provided hands-on experience that allowed students to apply theoretical knowledge, develop strategies, and make decisions in a risk-free environment. This hands-on experience is essential for a deeper understanding of the challenges and solutions of cloud migration projects. This was echoed by student S10, who noted that "this exercise also reflects the real world in which cloud consultants work". However, several students also voiced criticism, claiming "the chat itself was not interactive enough" and that "a real conversation with a supervisor would be more emotional and interactive and less bulleted."

Overall the feedback from the simulation was overwhelmingly positive and the role of ChatGPT in the simulation was found to be enriching. The students appreciated ChatGPT as an immediate source of information and assistance, which made the simulation more enriching. This enhanced their learning experiences while also making the simulation more realistic. Students enjoyed the simulation, as evidenced by the following statement from student S6: "This negotiation session was very interesting, exciting, and extremely realistic. Interacting with ChatGPT also promoted the development of critical thinking skills, as students had to learn to assess the relevance and accuracy of the information they received. Students highlighted the importance of this type of hands-on experience in vocational education, especially in rapidly evolving fields, such as cloud technology. They saw the simulation as an effective way to reinforce learning, develop problem-solving skills and encourage teamwork. They believed that the simulation was a useful tool for fostering teamwork, strengthening problem-solving abilities, and reinforcing learning.

## 4.2 ChatGPT

The 'ChatGPT' category reflects the students' direct experiences and evaluations of using ChatGPT as a tool in the simulated cloud migration consultancy sessions. These reflections include assessments of the usefulness, accuracy and limitations of ChatGPT in a professional consultancy context.

Students described their interaction with ChatGPT as eye-opening in terms of the potential of artificial intelligence to aid decision making. They highlighted ChatGPT's ability to provide comprehensive information quickly, which can be invaluable in guidance situations. The speed and efficiency with which ChatGPT responded to queries was highlighted as particularly helpful in accelerating the learning process and improving decision-making. This is evidenced by student S16's statement that "the accuracy of the questions asked by ChatGPT [...] was impressive" and student S20's statement that "the way the chatbot behaved was remarkably human-like".

However, challenges and limitations are also discussed. Some students pointed out that ChatGPTs' responses sometimes lacked depth or contextual understanding. This critical perception is illustrated by student S15, who felt that "ChatGPT is not mature enough to be a good negotiation partner". These limitations led to discussions about the need to be critical of the information provided and to evaluate it in the context of their own knowledge and the specific requirements of a cloud migration project.

Students also recognised the importance of developing skills in using AI tools. They reflected on the need to formulate clear and precise questions in order to get useful answers and saw this as an important skill for the future. Moreover, ChatGPT's function was emphasised as an addition to human knowledge rather than a substitute. Students saw ChatGPT as a tool that complements human skills by providing access to information that supports decision making without undermining critical judgment and human intuition. This collaborative approach fosters a synergy between AI and human capabilities, contributing to a more comprehensive and effective decision-making process.

### 4.3 Learning Experience

The Learning Experience category includes student reflections on their personal and professional growth processes while practising with ChatGPT in simulated cloud migration consulting scenarios. These reflections include insights into the effectiveness of ChatGPT as a learning tool, the exploration of students' learning strategies, and the development of cloud migration skills. This is clearly illustrated by the statement of student S5: "In conclusion, I can say with conviction that this role-play has fundamentally changed my perspective on cloud migration projects. It has given me the tools and understanding to tackle such projects with confidence and competence". In addition, students reported a deeper understanding of the complexities and challenges associated with cloud migration projects. They highlighted how interacting with ChatGPT enabled them to apply theoretical knowledge to practical scenarios, giving them a better understanding of how to plan and execute such projects. This is supported by student S6's comment: "It was quite fun to be able to draw on the knowledge I had already learned from the lectures, because these were the kinds of questions that were asked at the end." The opportunity to experiment with various approaches in a risk-free environment and to receive immediate feedback were also emphasised by the students as being crucial for encouraging deeper reflection and understanding.

Another important aspect of the learning experience is the improvement in communication skills. Students found that the need to articulate their thoughts and strategies clearly and concisely to interact effectively with ChatGPT improved their ability to communicate complex ideas clearly. This is particularly relevant in counselling situations where clear communication is crucial. Students also reflected on the role of AI in their education and professional development. Students also stressed the potential and limitations of AI-based learning tools, particularly in terms of personalised learning and adaptation to individual learning styles. The ChatGPT exercise is seen as an example of how AI can support learning by providing interactive and adaptive learning experiences.

To conclude students discussed the value of lifelong learning in the rapidly advancing field of technology. Many of them found that using ChatGPT highlighted the necessity to continuously learn new skills, but also to deepen and refine acquired knowledge to keep pace with developments in cloud computing and other fields. To sum up, the students "would recommend this method of learning to others" and "agree to expand on this exercise going forward."

## 5. DISCUSSION

The integration of artificial intelligence into university teaching using ChatGPT in business games demonstrates the potential of AI to revolutionise the educational landscape. The study, conducted as part of the Cloud Computing Engineering Masters programme, shows how AI may increase student engagement and motivation while also fostering the development of critical thinking, problem-solving, and effective communication skills. These findings resonate with Salmon [3], who emphasises the value of fusing the best of technology with the best of human learning. The targeted use of ChatGPT in a simulated role-playing game for the strategic planning and implementation of cloud migration enabled students to apply theoretical knowledge in a practical and risk-free environment. This game-based learning approach aligns with the research of Matute Vallejo and Melero [18], who advocate for simulation games as an effective method for improving learning engagement and performance. The positive feedback from students regarding their interaction with ChatGPT confirms the importance of AI as an innovative teaching tool that enriches the learning experience. This result echoes Schmid et al.'s [12], assertion about the potential of AI tools to create more flexible and personalised teaching and learning environments.

The study's outcomes underscore the transformative potential of AI in education, aligning with the European Commission's [6], call for the optimal use of AI applications in education to improve both learning outcomes and adaptability to digital change. In conclusion, the effective integration of AI tools, particularly ChatGPT, in simulation games actively engages students in the learning process and prepares them for the demands of the digitised work environment. These findings emphasise the necessity for continuous adaptation and innovation in higher education, as recommended by Tran and Herzig [4], advocating for increased integration of AI in higher education to ensure forward-looking and high-quality education.

In our study, we extend research on large language models (LLMs) such as ChatGPT by integrating them into role-based simulation games in an educational context. This application demonstrates the practical implementation of Vygotsky's sociocultural theory [14], by using ChatGPT to operationalise the Zone of Proximal Development (ZPD) and simulate social interactions. Our findings illustrate how digitised teaching methods enhance cognitive skills and offer new insights into the application of sociocultural practices in modern education.

Furthermore, our research supports the transformative effects of integrating AI into curricula, as discussed by Tran and Herzig [4], and Schmid et al. [12]. By using ChatGPT in role-playing games, we demonstrate how AI can enhance critical thinking and problem-solving skills. In addition, we address the need for better management tools for educators and the creation of personalised learning experiences, as highlighted by Hedderich et al. [32], Sonderegger and Seufert [33], and Gan et al. [34], emphasising the practical aspects of using AI in education.

## 6. LIMITATIONS AND FUTURE DIRECTIONS

This study adopts innovative approaches and provides insightful findings. However, it is important to consider a number of limitations when interpreting the results, most notably the fact that the findings are limited in their generalisability due to the focus on cloud computing engineering students which is a restricted group of 20 participants. While the exclusive use of qualitative data provides in-depth understanding of the learning experience, it neglects quantitative evaluations of learning success and engagement, which are essential for a comprehensive assessment of the effectiveness of ChatGPT in educational scenarios. In addition, the ethical aspects of the use of AI in education have not been fully addressed. Finally, the independent writing of reflections by students, without the possibility of consultation, limited the depth of the findings. The qualitative research approach could be seen as limiting from a quantitative perspective. Future research should therefore aim for a more diverse group of participants and methodological triangulation to enable a more comprehensive evaluation.

Expanding participant pools and using standardised questionnaires in conjunction with qualitative research approaches have to be the top priorities for future studies.This approach will increase the robustness and reliability of feedback collection and provide a more comprehensive understanding of how students interact with and benefit from these simulations. Using larger and more diverse groups of participants will allow for a broader assessment of educational impact across different learning environments and student demographics.

Additionally, it is critical to create resources that allow teachers to accurately manage and promote chatbot interactions in role-playing games. Tailoring chatbot behaviour to individual student responses and engagement can significantly enhance the learning experience, especially in complex or sensitive scenario simulations. Future studies should also focus on personalising the learning experience within these role-playing scenarios and integrating emotional intelligence into chatbots to make them more responsive to students' emotions. Establishing rigorous methods to evaluate the effectiveness of these technologies in role-playing simulations will help educators better understand their impact and refine their strategies to maximise educational benefits.

Despite the significant potential that Large Language Models (LLMs) such as ChatGPT offer for education, there are important limitations that need to be considered. The ethical concerns and privacy issues associated with the use of these technologies in educational contexts require careful consideration. It is crucial to protect sensitive student data and ensure privacy, which calls for well-designed regulation. Furthermore, our study did not fully assess the resources required to develop and maintain educational chatbots, which could affect the feasibility and sustainability of their implementation in educational institutions.

In addition, our study focuses primarily on the implications within the education sector and neglects the broader applications of AI in other sectors, potentially obscuring the full impact of these technologies. Ensuring that chatbots stay on topic throughout pre-defined dialogues presents additional challenges in addition to the ongoing demand for monitoring and adjustments. Furthermore, the issue of data bias in AI training, which can undermine the fairness and reliability of learning outcomes, has not been sufficiently addressed. Finally, the rapid evolution of AI technology may not have been fully considered in our analysis.

# 7. CONCLUSION

This study highlights the roles of AI and digital tools in modern education, particularly in the context of higher education. The integration of AI into teaching, as demonstrated by the use of ChatGPT in simulation games, not only offers students the opportunity to apply theoretical knowledge in practice, but also promotes important skills such as critical thinking, problem-solving and effective communication. These findings highlight the importance of further developing digital teaching methods to improve the quality of higher education and prepare both learners and teachers for the challenges and opportunities of the digitalised future. This practical case study shows that AI tools can be used not only for writing or research, but also for other applications in teaching. Even though Schöllhammer [35], argued against requiring students to use AI tools in the classroom, AI is an inevitable subject from a computer science standpoint for the Master's course in Cloud Computing Engineering. AI is therefore not only the content of the Master's programme, but is also used in a targeted didactic way. It should be noted that this case study is the first attempt at such a task. This particular example ought to inspire other university lecturers to reorganise their courses and experiment with different tasks. It is important to stress that this example should not be seen as complete or exhaustive. Rather, it is a first step that will be continuously improved and expanded in the coming years.